\documentclass[amsmath,amssymb,aps,prl,manuscript,superscriptaddress,showpacs,floatfix]{revtex4-1}

\usepackage{graphicx}
\usepackage[dvips]{color}
\usepackage{epstopdf}

\begin{document}

\title{Single-molecule enhanced spin-flip detection}

\author{Maider Ormaza}
\email{ormaza@ipcms.unistra.fr}
\author{Nicolas Bachellier}
\affiliation{IPCMS, CNRS UMR 7504, Universit\'{e} de Strasbourg, 67034 Strasbourg, France}
\author{Marisa N. Faraggi}
\affiliation{Ecole Normale Sup\'erieure, D\'epartement de Chimie, ENS-CNRS-UPMC UMR 8640, 75005 Paris, France}
\author{Benjamin Verlhac}
\affiliation{IPCMS, CNRS UMR 7504, Universit\'{e} de Strasbourg, 67034 Strasbourg, France}
\author{Paula Abufager}
\affiliation{Instituto de F\'{i}sica de Rosario, Consejo Nacional de Investigaciones Cient\'{i}ficas y T\'ecnicas (CONICET) and Universidad Nacional de Rosario, Av. Pellegrini 250 (2000) Rosario, Argentina}
\author{Philippe Ohresser}
\affiliation{Synchrotron SOLEIL, L'Orme des Merisiers, Saint-Aubin - BP 48, 91192 Gif-sur-Yvette, France}
\author{Lo{\"i}c Joly}
\author{Michelangelo Romeo}
\author{Fabrice Scheurer}
\affiliation{IPCMS, CNRS UMR 7504, Universit\'{e} de Strasbourg, 67034 Strasbourg, France}
\author{Marie-Laure Bocquet}
\affiliation{Ecole Normale Sup\'erieure, D\'epartement de Chimie, ENS-CNRS-UPMC UMR 8640, 75005 Paris, France}
\author{Nicol\'as Lorente}
\affiliation{Centro de F{\'{\i}}sica de Materiales CFM/MPC (CSIC-UPV/EHU), Paseo Manuel de Lardizabal 5, 20018 Donostia-San Sebasti\'an, Spain}
\affiliation{Donostia International Physics Center (DIPC), Paseo Manuel de Lardizabal 4, E-20018 Donostia-San Sebasti\'an, Spain}
\author{Laurent Limot}
\email{limot@ipcms.unistra.fr}
\affiliation{IPCMS, CNRS UMR 7504, Universit\'{e} de Strasbourg, 67034 Strasbourg, France}

\date{\today}

\begin{abstract}
We studied the spin-flip excitations of a double-decker nickelocene molecule (Nc) adsorbed on Cu(100) by means of inelastic tunneling spectroscopy (IETS), X-ray magnetic circular dichroism (XMCD) and density functional theory calculations (DFT). The results show that the molecule preserves its magnetic moment and magnetic anisotropy not only on Cu(100), but also in different metallic environments including the tip apex. Taking advantage of the efficient spin-flip excitation of this molecule, we show how such a molecular functionalized tip boosts the inelastic signal of a surface supported Nc by almost one order of magnitude thanks to a double spin-excitation process. 
\end{abstract}

\maketitle 


Recent advances in addressing and controlling the spin states of a surface-supported object --atom or molecule-- have further accredited the prospect of quantum computing~\cite{molecularnanomagnets,Loss2001} and of an ultimate data-storage capacity~\cite{Gambardella2003}. Information encoding requires that the object must possess stable magnetic states, in particular magnetic anisotropy to yield distinct spin-dependent states in the absence of a magnetic field together with long magnetic relaxation times~\cite{Zyazin2010,Thiele2014, Rau2014,Donati2016}. Scanning probe techniques have shown that inelastic electron tunneling spectroscopy (IETS) within the junction of a scanning tunneling microscope (STM) is a good starting point to study the stability of these spin states. STM-IETS allows for an all-electrical characterization of these states by promoting and detecting spin-flip excitations within the object of interest~\cite{Heinrich2004,Hirjibehedin2007,Chen2008,Tsukahara2009,Khajetoorians2015,Jacobson2015}. It can also provide an electrical control over these states~\cite{Loth2010b}, simplifying the information readout process. As spin excitations need however to be preserved from scattering events with itinerant electrons, single objects are usually placed on non-metallic surfaces such as thin-insulating layers~\cite{Loth2010,Rau2014,Yan2015,Donati2016} or superconductors~\cite{Heinrich2013}.

In this sense, new approaches to improve the detection of spin-flip excitations are desirable. With this purpose we present here a novel strategy based on the molecular functionalization of a STM tip. Atomic point-like probe-particles, such as  CO, H$_2$, D$_2$, Xe or CH$_4$, decorating scanning probe tips have been proved to enhance the imaging resolution to the extent of resolving the internal structure of large organic molecules, to act as force sensors and signal transducers, to image the local potential energy landscape of an adsorbed molecule, or to improve the detection of molecular vibrations~\cite{Wagner2015, Temirov2008, Weiss2010, Gross2011, Ho2014, Kichin2011, Kichin2013, Hapala2014, Gross2009, Weiss2010, Guo2016}. Tip functionalization with larger organic molecules could introduce additional mechanical, electronical or/and spin degrees of freedom very valuable for further development of the scanning probe techniques capabilities~\cite{Wagner2015b}.

\begin{figure}
  \begin{center}
    \includegraphics[width=0.55\textwidth]{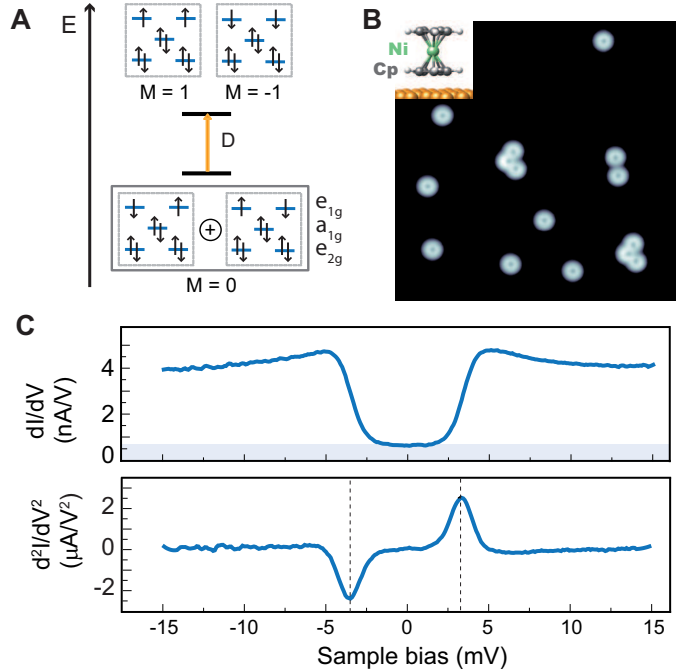}
  \end{center}
  \caption{\textbf{Spin-flip excitations of a single Nc molecule.} (a) Energy diagram of the zero-field splitting for Nc. The ground state is $S$=1, $M$=0 and the excited states correspond to $S$=1, $M$=$\pm$1. (b) STM topography ($14.5\times14.5$~nm$^2$, $-15$~mV, $20$~pa) of Nc molecules adsorbed on a Cu(100) surface. Inset: Nc structure calculated by DFT. White, grey and green atoms correspond to H, C and Ni atoms. (c) Differential conductance spectrum and its derivative measured at $2.4$~K acquired at $50$~pA and $-15$~mV. The spectrum was measured by means of a lock-in amplifier using a modulation of 150~$\mu$V rms at $722$~Hz. The $d^2I/dV^2$ versus $V$ spectrum was obtained by numerically deriving the $dI/dV$ spectrum.
}
\label{figure1}
\end{figure}

In this work, we are able to controllably attach a double-decker molecule, nickelocene (Nc hereafter), to the tip apex and exploit the spin-flip excitations exhibited by Nc to boost the signal of an IETS measurement. The chemical structure of nickelocene consists in a Ni atom sandwiched between two cyclopentadienyl rings (C$_5$H$_5$, Cp hereafter), endowing nickelocene with an electronic structure (a$_{1\text{g}}$)$^2$(e$_{2\text{g}}$)$^4$(e$_{1\text{g}}$)$^2$ of spin $S=1$ in the gas phase (Fig.\,\ref{figure1}a)~\cite{Feng2003}. By attaching a nickelocene to the tip apex of a STM, we show that two spin-flip excitations can be sequentially triggered, resulting in a large enhancement of the inelastic signal as illustrated by the IETS acquired above a surface-supported Nc. This unique situation can be achieved due to the chemical stability, the presence of magnetic anisotropy, the full preservation of the Nc spin on the tip apex, or more generally in a metallic environment, and the extremely efficient spin inelastic effect exhibited by this molecule.

\begin{figure}
  \begin{center}
    \includegraphics[width=0.55\textwidth]{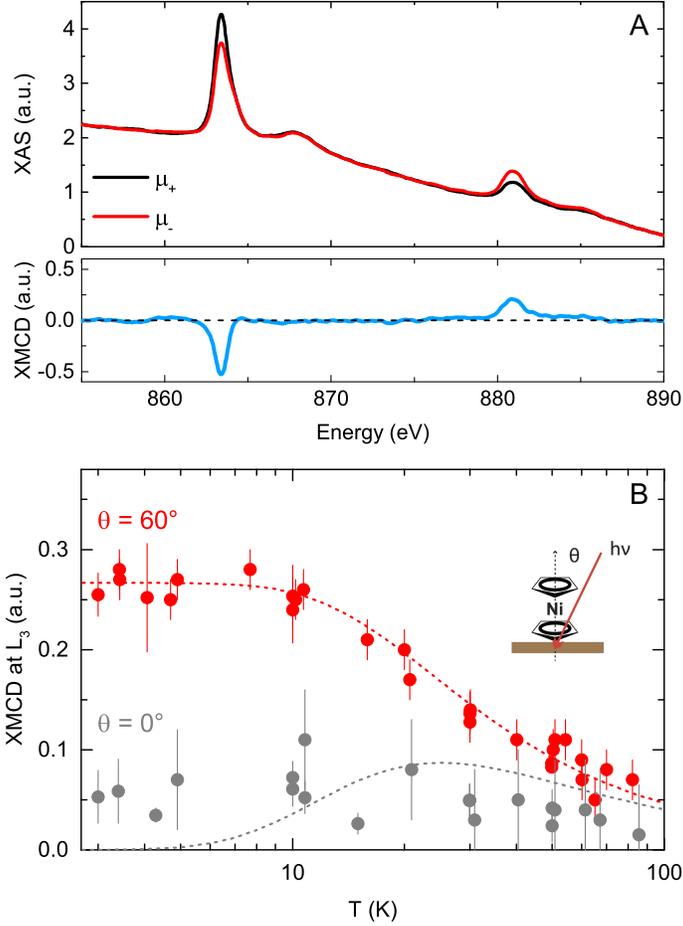}
  \end{center}
  \caption{\textbf{XMCD measurements.} (a) Absorption spectra for left and right circularly polarized light and XMCD signal  at the Ni $L_{2,3}$ edges in a 6.5\,T magnetic field (60$^{\circ}$ off-normal) at 4.7\,K for a collection of isolated molecules. (b) Maximum value of the $L_{3}$ XMCD signal for normal (grey circle) and 60$^{\circ}$  (red circle) off-normal incidence under a 6.5\,T field as a function of temperature. The dashed lines correspond to simulations for a $S=1$ system with a magnetic anisotropy of $D=3.2$~meV; the simulations are scaled to the data by a factor $0.8$. Note that the magnetic field is parallel to the X-ray beam.}
\label{figure2}
\end{figure}

To benchmark the electronic and magnetic properties of a Nc-terminated tip, we investigated first the nickelocene adsorption on a Cu(100) surface using a metal tip. The ring-shaped pattern visible in Fig.~\ref{figure1}b indicates that individual nickelocene molecules bind to the surface through a Cp ring, while the other Cp ring is exposed to vacuum (see experimental details in supplementary material~\cite{SI}). In particular, the Cp ring is centered on a hollow position of Cu(100) most likely adopting the $D_{5h}$ symmetry (eclipsed configuration)~\cite{Bachellier2016}. Figure~\ref{figure1}c presents the differential conductance ($dI/dV$) versus bias ($V$) spectrum of an isolated Nc molecule by positioning the tip above the Cp ring, together with its numerical derivative. As shown, the $dI/dV$ spectrum is reduced near zero bias (Fig.~\ref{figure1}c) with symmetric steps up to 600$\%$ times higher in conductance at an energy of $\lvert3.2\pm0.1\rvert$~meV (the uncertainty translates the tip-dependency observed). Since the vibrational excitations of nickelocene adsorbed on the surface can be \textit{a priori} discarded in this energy range~\cite{Pugmire2001}, the differential conductance steps reveal the occurrence of spin-flip excitations within the molecule. In view of the spin $1$ character predicted upon adsorption on Cu(100)~\cite{Bachellier2016}, these excitations occur between the ground state ($S$=1, $M$=0) and the doubly-degenerated ($S$=1, $M$=$\pm$1) excited states of the molecule. The threshold energy observed in IETS corresponds then to a direct measurement of the longitudinal magnetic anisotropy energy, $D$~\cite{SI}. The large step height observed in the conductance reflects a highly efficient spin-flip inelastic effect which we attribute to the interference of two spin-channels with different symmetries, an effect not seen in previous experiments where only one channel prevails~\cite{SI}.

To further characterize the magnetic origin of the excitation, we performed X-ray magnetic circular dichroism (XMCD) measurements in a magnetic field of $6.5$~T on a collection of individual Nc molecules on Cu(100)~\cite{SI}. The top panel of Fig.~\ref{figure2}a shows X-ray adsorption spectra (XAS) recorded at the $L_{2,3}$ edges of Ni for left and right circularly polarized light. The panel below presents the difference between them, the so-called XMCD spectrum. Having a clear dichroic signal is already an indication of the magnetic character of nickelocene on Cu(100). Figure~\ref{figure2}b shows the evolution of the signal maximum at the Ni $L_{3}$ edge as a function of the temperature for grazing (red dots, $\theta=60^{\circ}$) and normal (black dots, $\theta=0^{\circ}$) incidence of the applied magnetic field. At low temperature, the signal shows a clear angular dependence revealing the existence of an easy magnetization direction, more specifically of an easy plane perpendicular to the molecular axis. For the sake of comparison, we have simulated~\cite{SI} in Fig.~\ref{figure2}b the magnetization of a $S=1$ system under an external field of $6.5$~T and possessing a magnetic anisotropy of $D=3.2$~meV. Based on the XMCD results, we can assert that $D$ is positive, implying that $M=0$ corresponds to the ground state of nickelocene and the $M=\pm 1$ to the excited state.

\begin{figure}
  \begin{center}
    \includegraphics[width=0.7\textwidth]{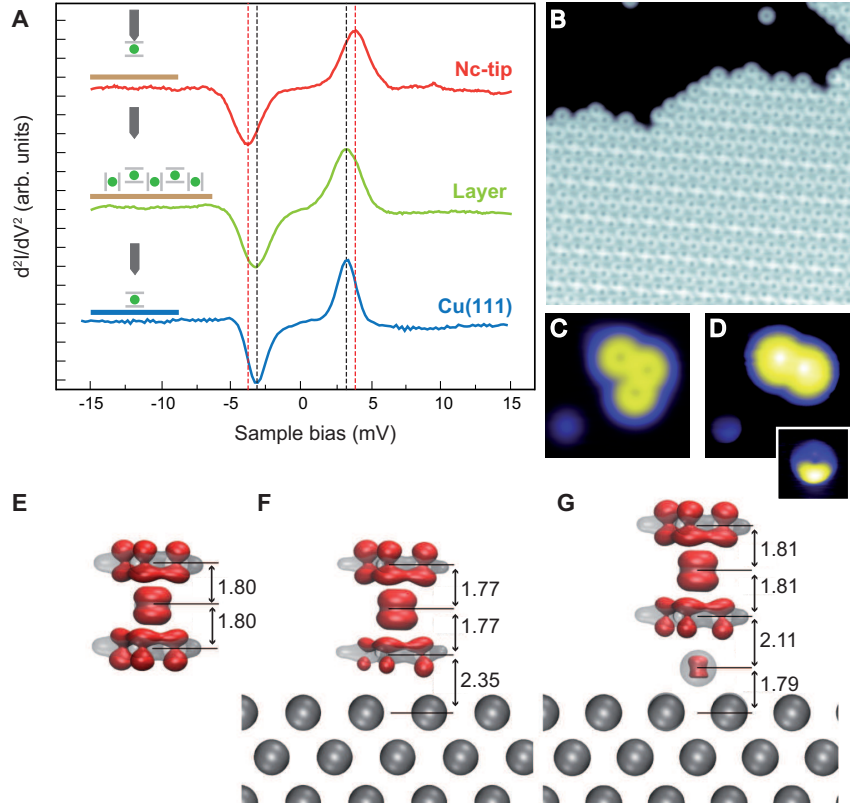}
  \end{center}
  \caption{\textbf{Nc molecule in different environments.} (a) Second derivative of the differential conductance measured for three different environments: isolated Nc adsorbed on Cu(111) (in blue), Nc within the molecular self-assembled monolayer on Cu(100) (in green) and Nc adsorbed on the tip (in red). The spectra were taken with different feedback parameters and overlaid to highlight the different bias thresholds. (b) STM topography of the self-assembled Nc molecules forming a compact arrangement on Cu(100). Images (c) before and (d) after the transfer of Nc to the tip. The protrusion on the left corresponds to a Cu atom. With the Nc-terminated tip, the Cu atom exhibits a molecular pattern as visible in the inset of panel (d). (e-g) Spin density (majority spin in red, negligible minority spin) and total electron density (grey transparent) of gas phase Nc with MAE= $-1.54$~meV (e), adsorbed Nc on Cu(100) with MAE= $-1.27$~meV (f) and adsorbed Cu-Nc on Cu(100) with MAE= $-1.49$~meV (g). A negative MAE corresponds to an easy plane, in other words to a positive value of $D$. The isosurface of the spin (total electron) density  is 0.02 (1.3) electron per {\AA}$^3$. Vertical heights are given in {\AA}. Image parameters: (b) $18\times18$~nm$^2$, -20\,mV, 20\,pA, (c) and (d) $3.5\times3.5$~nm$^2$ (Inset of panel D: $1.5\times1.5$~nm$^2$). }
\label{CuNc}
\end{figure}

The magnetic properties of isolated nickelocene turn out to be robust to external perturbations, possibly owing this behavior to the cyclic $\pi$* orbital of the Cp rings sandwiching the Ni atom. This differentiates Nc from other systems reported so far~\cite{Hirjibehedin2007,Miyamachi2013,Gambardella2003,Donati2016} in which the local environment is a crucial factor determining the magnetic anisotropy and/or the spin. At higher coverages, for example, where Nc molecules self-assemble following a T-shaped geometry (Fig.~\ref{CuNc}b) typical for metallocenes~\cite{Ormaza2015,Bachellier2016}, no difference was observed in the IETS spectra of upstanding molecules within the layer (Fig.~\ref{CuNc}a). We have checked the impact of the surface symmetry by changing to Cu(111), also resulting in no difference for isolated or self-assembled nickelocene (Fig.~\ref{CuNc}a). Supporting these findings, we note that the spin $1$ character and magnetic anisotropy are consistent with existing macroscopic measurements of Nc powder samples and crystallites ~\cite{Prins1967,Baltzer1988} or matrix embedded samples~\cite{Li1992} where $D$ ranges from $3.17$~meV to $4.16$~meV; recent theoretical calculations for gas-phase Nc also predict $D$ to be close to 5~meV~\cite{Vaara2015}. In the following, we exploit such robustness to produce a portable source for spin-flip excitations by attaching a single nickelocene to the tip apex of the STM.

\begin{figure}
  \begin{center}
    \includegraphics[width=0.7\textwidth]{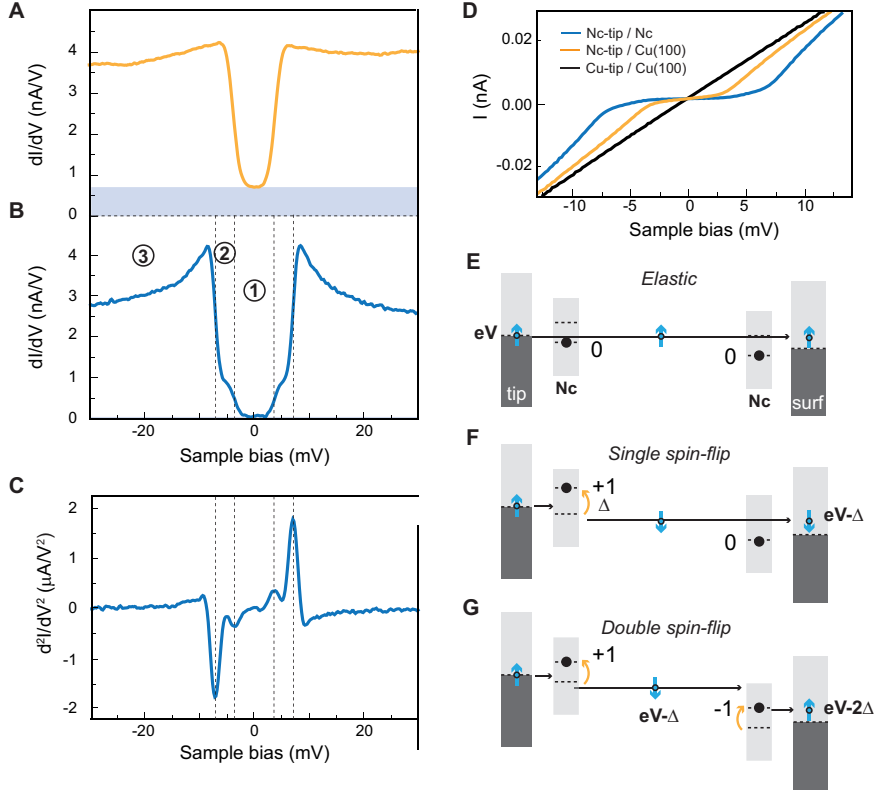}
  \end{center}
  \caption{\textbf{Double spin-flip excitation.} (a) $dI/dV$ spectrum of a Nc-terminated tip above Cu(100). Panel (b) and (c) show the $dI/dV$ spectrum of a Nc molecule measured with a Nc-terminated tip and its derivative. Three tunneling regimes are highlighted (see text) indicated by regions 1, 2 and 3. To record the spectra the feedback loop was opened at 100\,pA and -40\,mV. Note that the different elastic background observed in (a) and (b) is due to a larger tip-surface distance in (b). Panel (d) presents the $I(V)$ curve showing the opening of the inelastic channels. Panels (e) to (g) present a sketch of the different tunneling mechanisms: (e) there is only elastic tunneling of electrons; (f) the spin of just one of the molecules is excited. In this case the excitation of the molecule on the tip from $M=0$ to $M=+1$ is shown, (g) both molecules undergo spin-flip excitations.}
\label{Nctip}
\end{figure}

Similarly to Xe atoms~\cite{Eigler1991}, or physisorbed molecules~\cite{Bartels1998,Ormaza2016}, we can transfer a Nc to our monoatomic tip apex by scanning the target molecule at biases close to $-1$~mV (Figs.~\ref{CuNc}c and \ref{CuNc}d). The tip functionalization is reversible, i.e., in the same way the molecule goes to the tip it can be released back to an atom on the surface ($+1$~mV), indicating that the molecule remains intact during the transfer process. Counter-images of the molecular tip acquired above a Cu atom (inset of Fig.~\ref{CuNc}d) confirm the presence of a nickelocene molecule. The half-moon shape that is observed can be easily assigned to a tilted Nc~\cite{Heinrich2011}. The nickelocene on the tip exhibits similar magnetic properties as previously seen for nickelocene on the surface (Figs.~\ref{CuNc}a and ~\ref{Nctip}a). IETS acquired above the bare surface reveals a larger magnetic anisotropy of $(D=3.7\pm0.3)$~meV and conductance steps up to 900$\%$ times higher compared to the elastic conductance alone. 

The magnetic properties of the molecular tip are computed by fully-relaxing a single nickelocene atop a Cu atom adsorbed on Cu(100). In Figures \ref{CuNc}e-g, the spin density of Nc in three different environments --gas phase, adsorbed on Cu(100) and adsorbed on a Cu atom on Cu(100)-- is shown together with the total electron density. The spin density is similar in the three cases showing that the molecule keeps a spin equal to $1$. The spin density follows the density of the two degenerate frontier molecular orbitals, with weight on the $d_{xz}$ and $d_{yz}$ orbitals of the Ni atom and the $\pi$* orbital of the Cps. For the Cu atom below Nc in Fig.~\ref{CuNc}f, there is a very small spin polarization that accounts for 2$\%$ of the total polarization. From both adsorbed cases, our calculations show that the surface polarizes with a minimum impact on the molecule and the molecular spin is preserved in the three systems studied, obtaining a negligible value for the orbital moment of $0.1\mu_\text{B}$. An inspection of the molecular geometry shows a small reduction of the Nc height (3.54~\AA) upon adsorption on Cu(100) due to the depopulation of the anti-bonding frontier molecular orbital. This corresponds to a very small (0.1 $e^-$) electron transfer from the molecule to the surface. Interestingly, the charge transfer is reversed when Nc is on the Cu atom; the molecule charge increases by 0.14 $e^-$ leading to a larger molecular height (3.62~\AA). This translates into a different magnetic anisotropy energy (MAE)~\cite{SI}. The MAE is smallest when the molecule is on the surface (Fig.~\ref{CuNc}f, -1.27~meV) and recovers the gas phase value (Fig.~\ref{CuNc}e, -$1.54$~meV) when adsorbed on the Cu atom (Fig.~\ref{CuNc}g, -1.49~meV); the $17\%$ difference in the MAE is consistent with the experimental changes found for $D$ for Nc on the surface and on the tip. Note that the MAE is essentially affected by the local ligand field as remarked in other systems~\cite{Heinrich2015}.

The presence of Nc at the tip apex offers the thrilling perspective of an enhanced detection of on-surface spin-flip excitations. To exemplify this effect we acquired IETS spectra above of a Nc molecule adsorbed on the surface using a molecular tip (see Fig.~\ref{Nctip}a-c). In this case, along with the conductance steps at $\Delta=\lvert3.4\pm0.2\rvert$~meV, a new set of prominent steps appears at $\lvert6.9\pm0.4\rvert$~meV, which are the signature of a double spin-excitation process sequentially involving the nickelocene on the tip and on the surface, or \textit{vice versa}. These inelastic processes can be directly detected in the $I(V)$ curve shown in Fig.\,\ref{Nctip}d. To explain the origin of these steps, we suppose first, for simplicity, that the molecules on the tip and on the surface have the same threshold energy $\Delta$. When a tunneling electron has an energy below $\Delta$ (region 1) only elastic tunneling processes can take place (Fig.\,\ref{Nctip}e). Above that threshold but below $2 \Delta$ (region 2), one of the molecular spins, either the one on the tip or the one on the surface, can be excited (Fig.\,\ref{Nctip}f). Above $2 \Delta$ (region 3) a tunneling electron has sufficient energy to excite inelastically both molecules (Fig.~\ref{Nctip}g). Taking now into account the slightly different threshold energies as evidenced above for Nc on the tip ($3.7$~meV) and on the surface ($3.2$~meV), we find that their sum nicely matches the threshold energy found for a double excitation. In principle, in region 2 two additional steps (at $3.2$~meV and $3.7$~meV) rather than one ($3.4$~meV) should be observed, but their detection is beyond our energy resolution. 

The most prominent effect associated to the double spin excitation is the drastic enhancement of the conduction step height observed at $2\Delta$ (a factor 43 with respect the elastic signal in the spectrum of Fig.~\ref{Nctip}b) that we explain as follows. Let us suppose that the conductance is proportional to the probability of taking one electron from one electrode (tip) into the other electrode (surface), in other words that the conductance is proportional to the $T$-matrix modulus squared~\cite{SI}. Since the $T$-matrix has now spin degrees of freedom in the tip and on the sample, we denote the new matrix elements as $T_{n,m}$ where $n$ is the number of excitations of the tip and $m$ the number of excitations of the sample. Then, the conductance will be proportional to $\sum_{n,m} \vert T_{n,m} \vert^2$, with a  sum on the excitations since the channels are supposed independent. In this case, $T_{0,0}$ corresponds to the elastic transmission of an electron from the tip to the surface or \textit{vice versa} (region 1). $T_{1,0}$ and $T_{0,1}$ correspond to the case in which there is only one inelastic event during the process, i.e., in the molecule on the tip, or on the surface. Finally, $T_{1,1}$ is related to the probability of an electron being transmitted inelastically between the electrodes exciting both molecules. Hence, if the applied bias is high enough to open the first excitation channel for the molecules on the tip and on the sample (region 3), we obtain that the conductance is proportional to $\vert T_{0,0} \vert^2+\vert T_{1,0} \vert^2+ \vert T_{0,1} \vert ^2+ \vert T_{1,1} \vert^2$. The observed inelastic \textit{versus} elastic ratios show tip dependency.  Considering the maximum step height values we observed, we have for the Nc on the tip that $(\vert T_{1,0}\vert^2 +\vert T_{0,0}\vert^2)      /\vert T_{0,0}\vert^2\approx 9$ and for the Nc on the surface $(\vert T_{0,1} \vert ^2+ \vert T_{0,0}\vert^2)/ \vert T_{0,0}\vert^2 \approx  6$. Since the excitations of the molecule on the tip and on the sample are independent, the transmission probability of a double spin-flip excitation to occur is approximately $\vert T_{1,0}\vert^2 \vert T_{0,1}\vert^2  /\vert T_{0,0}\vert^4 =\vert T_{1,1}\vert^2/\vert T_{0,0}\vert^2\approx 8\cdot 5=40$. The conductance is then in region 3,  $(\vert T_{0,0} \vert^2+\vert T_{1,0} \vert^2+ \vert T_{0,1} \vert ^2+ \vert T_{1,1} \vert^2)/\vert T_{0,0} \vert^2$=1+6+9+40, this is 56 times higher than the elastic conductance alone (region 1).

To summarize, we have shown that a double molecular spin-flip excitation leading to an enhancement of the inelastic signal in spin-flip spectroscopy can be obtained using a Nc-terminated tip. The tip-functionalization scheme described can be generalized to study the spin-flip excitations of other systems~\cite{Chen2008, Tsukahara2009}, provided that the inelastic energy thresholds are sufficiently large as to be resolved by the given experimental resolution. The enhanced detection will be in any case favored when the inelastic channel transmission is larger than the elastic one. On a more fundamental note, the portability of the spin-flip excitations, whether of nickelocene or eventually of other molecules, could serve as a spin-sensitive probe (with or without magnetic field) and thereby provide quantitative spin-dependent STM data.

\begin{acknowledgments}
This work has been supported by the Agence Nationale de la Recherche (Grant No. ANR-13-BS10-0016, ANR-11-LABX-0058 NIE, ANR-10-LABX-0026 CSc). Experiments were performed on the DEIMOS beamline\cite{deimos} at SOLEIL Synchrotron, France (proposal number 20150309). We are grateful to the SOLEIL staff for smoothly running the facility. We thank the national computational center IDRIS, CINES and TGCC (Grant - [x2015087364]) for CPU time. 
\end{acknowledgments}


\begin{thebibliography}{46}%
\makeatletter
\providecommand \@ifxundefined [1]{%
 \@ifx{#1\undefined}
}%
\providecommand \@ifnum [1]{%
 \ifnum #1\expandafter \@firstoftwo
 \else \expandafter \@secondoftwo
 \fi
}%
\providecommand \@ifx [1]{%
 \ifx #1\expandafter \@firstoftwo
 \else \expandafter \@secondoftwo
 \fi
}%
\providecommand \natexlab [1]{#1}%
\providecommand \enquote  [1]{``#1''}%
\providecommand \bibnamefont  [1]{#1}%
\providecommand \bibfnamefont [1]{#1}%
\providecommand \citenamefont [1]{#1}%
\providecommand \href@noop [0]{\@secondoftwo}%
\providecommand \href [0]{\begingroup \@sanitize@url \@href}%
\providecommand \@href[1]{\@@startlink{#1}\@@href}%
\providecommand \@@href[1]{\endgroup#1\@@endlink}%
\providecommand \@sanitize@url [0]{\catcode `\\12\catcode `\$12\catcode
  `\&12\catcode `\#12\catcode `\^12\catcode `\_12\catcode `\%12\relax}%
\providecommand \@@startlink[1]{}%
\providecommand \@@endlink[0]{}%
\providecommand \url  [0]{\begingroup\@sanitize@url \@url }%
\providecommand \@url [1]{\endgroup\@href {#1}{\urlprefix }}%
\providecommand \urlprefix  [0]{URL }%
\providecommand \Eprint [0]{\href }%
\providecommand \doibase [0]{http://dx.doi.org/}%
\providecommand \selectlanguage [0]{\@gobble}%
\providecommand \bibinfo  [0]{\@secondoftwo}%
\providecommand \bibfield  [0]{\@secondoftwo}%
\providecommand \translation [1]{[#1]}%
\providecommand \BibitemOpen [0]{}%
\providecommand \bibitemStop [0]{}%
\providecommand \bibitemNoStop [0]{.\EOS\space}%
\providecommand \EOS [0]{\spacefactor3000\relax}%
\providecommand \BibitemShut  [1]{\csname bibitem#1\endcsname}%
\let\auto@bib@innerbib\@empty
\bibitem [{\citenamefont {Gatteschi}\ \emph {et~al.}(2006)\citenamefont
  {Gatteschi}, \citenamefont {Sessoli},\ and\ \citenamefont
  {Villain}}]{molecularnanomagnets}%
  \BibitemOpen
  \bibfield  {author} {\bibinfo {author} {\bibfnamefont {D.}~\bibnamefont
  {Gatteschi}}, \bibinfo {author} {\bibfnamefont {R.}~\bibnamefont {Sessoli}},
  \ and\ \bibinfo {author} {\bibfnamefont {J.}~\bibnamefont {Villain}},\
  }\href@noop {} {\emph {\bibinfo {title} {Molecular Nanomagnets}}}\ (\bibinfo
  {publisher} {OUP Oxford},\ \bibinfo {year} {2006})\BibitemShut {NoStop}%
\bibitem [{\citenamefont {Leuenberger}\ and\ \citenamefont
  {Loss}(2001)}]{Loss2001}%
  \BibitemOpen
  \bibfield  {author} {\bibinfo {author} {\bibfnamefont {M.~N.}\ \bibnamefont
  {Leuenberger}}\ and\ \bibinfo {author} {\bibfnamefont {D.}~\bibnamefont
  {Loss}},\ }\href {http://dx.doi.org/10.1038/35071024} {\bibfield  {journal}
  {\bibinfo  {journal} {Nature}\ }\textbf {\bibinfo {volume} {410}},\ \bibinfo
  {pages} {789} (\bibinfo {year} {2001})}\BibitemShut {NoStop}%
\bibitem [{\citenamefont {Gambardella}\ \emph {et~al.}(2003)\citenamefont
  {Gambardella}, \citenamefont {Rusponi}, \citenamefont {Veronese},
  \citenamefont {Dhesi}, \citenamefont {Grazioli}, \citenamefont {Dallmeyer},
  \citenamefont {Cabria}, \citenamefont {Zeller}, \citenamefont {Dederichs},
  \citenamefont {Kern}, \citenamefont {Carbone},\ and\ \citenamefont
  {Brune}}]{Gambardella2003}%
  \BibitemOpen
  \bibfield  {author} {\bibinfo {author} {\bibfnamefont {P.}~\bibnamefont
  {Gambardella}}, \bibinfo {author} {\bibfnamefont {S.}~\bibnamefont
  {Rusponi}}, \bibinfo {author} {\bibfnamefont {M.}~\bibnamefont {Veronese}},
  \bibinfo {author} {\bibfnamefont {S.~S.}\ \bibnamefont {Dhesi}}, \bibinfo
  {author} {\bibfnamefont {C.}~\bibnamefont {Grazioli}}, \bibinfo {author}
  {\bibfnamefont {A.}~\bibnamefont {Dallmeyer}}, \bibinfo {author}
  {\bibfnamefont {I.}~\bibnamefont {Cabria}}, \bibinfo {author} {\bibfnamefont
  {R.}~\bibnamefont {Zeller}}, \bibinfo {author} {\bibfnamefont {P.~H.}\
  \bibnamefont {Dederichs}}, \bibinfo {author} {\bibfnamefont {K.}~\bibnamefont
  {Kern}}, \bibinfo {author} {\bibfnamefont {C.}~\bibnamefont {Carbone}}, \
  and\ \bibinfo {author} {\bibfnamefont {H.}~\bibnamefont {Brune}},\ }\href
  {\doibase 10.1126/science.1082857} {\bibfield  {journal} {\bibinfo  {journal}
  {Science}\ }\textbf {\bibinfo {volume} {300}},\ \bibinfo {pages} {1130}
  (\bibinfo {year} {2003})}\BibitemShut {NoStop}%
\bibitem [{\citenamefont {Zyazin}\ \emph {et~al.}(2010)\citenamefont {Zyazin},
  \citenamefont {van~den Berg}, \citenamefont {Osorio}, \citenamefont {van~der
  Zant}, \citenamefont {Konstantinidis}, \citenamefont {Leijnse}, \citenamefont
  {Wegewijs}, \citenamefont {May}, \citenamefont {Hofstetter}, \citenamefont
  {Danieli},\ and\ \citenamefont {Cornia}}]{Zyazin2010}%
  \BibitemOpen
  \bibfield  {author} {\bibinfo {author} {\bibfnamefont {A.~S.}\ \bibnamefont
  {Zyazin}}, \bibinfo {author} {\bibfnamefont {J.~W.~G.}\ \bibnamefont {van~den
  Berg}}, \bibinfo {author} {\bibfnamefont {E.~A.}\ \bibnamefont {Osorio}},
  \bibinfo {author} {\bibfnamefont {H.~S.~J.}\ \bibnamefont {van~der Zant}},
  \bibinfo {author} {\bibfnamefont {N.~P.}\ \bibnamefont {Konstantinidis}},
  \bibinfo {author} {\bibfnamefont {M.}~\bibnamefont {Leijnse}}, \bibinfo
  {author} {\bibfnamefont {M.~R.}\ \bibnamefont {Wegewijs}}, \bibinfo {author}
  {\bibfnamefont {F.}~\bibnamefont {May}}, \bibinfo {author} {\bibfnamefont
  {W.}~\bibnamefont {Hofstetter}}, \bibinfo {author} {\bibfnamefont
  {C.}~\bibnamefont {Danieli}}, \ and\ \bibinfo {author} {\bibfnamefont
  {A.}~\bibnamefont {Cornia}},\ }\href {\doibase 10.1021/nl1009603} {\bibfield
  {journal} {\bibinfo  {journal} {Nano Lett.}\ }\textbf {\bibinfo {volume}
  {10}},\ \bibinfo {pages} {3307} (\bibinfo {year} {2010})}\BibitemShut
  {NoStop}%
\bibitem [{\citenamefont {Thiele}\ \emph {et~al.}(2014)\citenamefont {Thiele},
  \citenamefont {Balestro}, \citenamefont {Ballou}, \citenamefont {Klyatskaya},
  \citenamefont {Ruben},\ and\ \citenamefont {Wernsdorfer}}]{Thiele2014}%
  \BibitemOpen
  \bibfield  {author} {\bibinfo {author} {\bibfnamefont {S.}~\bibnamefont
  {Thiele}}, \bibinfo {author} {\bibfnamefont {F.}~\bibnamefont {Balestro}},
  \bibinfo {author} {\bibfnamefont {R.}~\bibnamefont {Ballou}}, \bibinfo
  {author} {\bibfnamefont {S.}~\bibnamefont {Klyatskaya}}, \bibinfo {author}
  {\bibfnamefont {M.}~\bibnamefont {Ruben}}, \ and\ \bibinfo {author}
  {\bibfnamefont {W.}~\bibnamefont {Wernsdorfer}},\ }\href {\doibase
  10.1126/science.1249802} {\bibfield  {journal} {\bibinfo  {journal}
  {Science}\ }\textbf {\bibinfo {volume} {344}},\ \bibinfo {pages} {1135}
  (\bibinfo {year} {2014})}\BibitemShut {NoStop}%
\bibitem [{\citenamefont {Rau}\ \emph {et~al.}(2014)\citenamefont {Rau},
  \citenamefont {Baumann}, \citenamefont {Rusponi}, \citenamefont {Donati},
  \citenamefont {Stepanow}, \citenamefont {Gragnaniello}, \citenamefont
  {Dreiser}, \citenamefont {Piamonteze}, \citenamefont {Nolting}, \citenamefont
  {Gangopadhyay}, \citenamefont {Albertini}, \citenamefont {Macfarlane},
  \citenamefont {Lutz}, \citenamefont {Jones}, \citenamefont {Gambardella},
  \citenamefont {Heinrich},\ and\ \citenamefont {Brune}}]{Rau2014}%
  \BibitemOpen
  \bibfield  {author} {\bibinfo {author} {\bibfnamefont {I.~G.}\ \bibnamefont
  {Rau}}, \bibinfo {author} {\bibfnamefont {S.}~\bibnamefont {Baumann}},
  \bibinfo {author} {\bibfnamefont {S.}~\bibnamefont {Rusponi}}, \bibinfo
  {author} {\bibfnamefont {F.}~\bibnamefont {Donati}}, \bibinfo {author}
  {\bibfnamefont {S.}~\bibnamefont {Stepanow}}, \bibinfo {author}
  {\bibfnamefont {L.}~\bibnamefont {Gragnaniello}}, \bibinfo {author}
  {\bibfnamefont {J.}~\bibnamefont {Dreiser}}, \bibinfo {author} {\bibfnamefont
  {C.}~\bibnamefont {Piamonteze}}, \bibinfo {author} {\bibfnamefont
  {F.}~\bibnamefont {Nolting}}, \bibinfo {author} {\bibfnamefont
  {S.}~\bibnamefont {Gangopadhyay}}, \bibinfo {author} {\bibfnamefont {O.~R.}\
  \bibnamefont {Albertini}}, \bibinfo {author} {\bibfnamefont {R.~M.}\
  \bibnamefont {Macfarlane}}, \bibinfo {author} {\bibfnamefont {C.~P.}\
  \bibnamefont {Lutz}}, \bibinfo {author} {\bibfnamefont {B.~A.}\ \bibnamefont
  {Jones}}, \bibinfo {author} {\bibfnamefont {P.}~\bibnamefont {Gambardella}},
  \bibinfo {author} {\bibfnamefont {A.~J.}\ \bibnamefont {Heinrich}}, \ and\
  \bibinfo {author} {\bibfnamefont {H.}~\bibnamefont {Brune}},\ }\href
  {\doibase 10.1126/science.1252841} {\bibfield  {journal} {\bibinfo  {journal}
  {Science}\ }\textbf {\bibinfo {volume} {344}},\ \bibinfo {pages} {988}
  (\bibinfo {year} {2014})}\BibitemShut {NoStop}%
\bibitem [{\citenamefont {Donati}\ \emph {et~al.}(2016)\citenamefont {Donati},
  \citenamefont {Rusponi}, \citenamefont {Stepanow}, \citenamefont
  {W{\"a}ckerlin}, \citenamefont {Singha}, \citenamefont {Persichetti},
  \citenamefont {Baltic}, \citenamefont {Diller}, \citenamefont {Patthey},
  \citenamefont {Fernandes}, \citenamefont {Dreiser}, \citenamefont {{\v
  S}ljivan{\v c}anin}, \citenamefont {Kummer}, \citenamefont {Nistor},
  \citenamefont {Gambardella},\ and\ \citenamefont {Brune}}]{Donati2016}%
  \BibitemOpen
  \bibfield  {author} {\bibinfo {author} {\bibfnamefont {F.}~\bibnamefont
  {Donati}}, \bibinfo {author} {\bibfnamefont {S.}~\bibnamefont {Rusponi}},
  \bibinfo {author} {\bibfnamefont {S.}~\bibnamefont {Stepanow}}, \bibinfo
  {author} {\bibfnamefont {C.}~\bibnamefont {W{\"a}ckerlin}}, \bibinfo {author}
  {\bibfnamefont {A.}~\bibnamefont {Singha}}, \bibinfo {author} {\bibfnamefont
  {L.}~\bibnamefont {Persichetti}}, \bibinfo {author} {\bibfnamefont
  {R.}~\bibnamefont {Baltic}}, \bibinfo {author} {\bibfnamefont
  {K.}~\bibnamefont {Diller}}, \bibinfo {author} {\bibfnamefont
  {F.}~\bibnamefont {Patthey}}, \bibinfo {author} {\bibfnamefont
  {E.}~\bibnamefont {Fernandes}}, \bibinfo {author} {\bibfnamefont
  {J.}~\bibnamefont {Dreiser}}, \bibinfo {author} {\bibfnamefont {{\v
  Z}.}~\bibnamefont {{\v S}ljivan{\v c}anin}}, \bibinfo {author} {\bibfnamefont
  {K.}~\bibnamefont {Kummer}}, \bibinfo {author} {\bibfnamefont
  {C.}~\bibnamefont {Nistor}}, \bibinfo {author} {\bibfnamefont
  {P.}~\bibnamefont {Gambardella}}, \ and\ \bibinfo {author} {\bibfnamefont
  {H.}~\bibnamefont {Brune}},\ }\href {\doibase 10.1126/science.aad9898}
  {\bibfield  {journal} {\bibinfo  {journal} {Science}\ }\textbf {\bibinfo
  {volume} {352}},\ \bibinfo {pages} {318} (\bibinfo {year}
  {2016})}\BibitemShut {NoStop}%
\bibitem [{\citenamefont {Heinrich}\ \emph {et~al.}(2004)\citenamefont
  {Heinrich}, \citenamefont {Gupta}, \citenamefont {Lutz},\ and\ \citenamefont
  {Eigler}}]{Heinrich2004}%
  \BibitemOpen
  \bibfield  {author} {\bibinfo {author} {\bibfnamefont {A.~J.}\ \bibnamefont
  {Heinrich}}, \bibinfo {author} {\bibfnamefont {J.~A.}\ \bibnamefont {Gupta}},
  \bibinfo {author} {\bibfnamefont {C.~P.}\ \bibnamefont {Lutz}}, \ and\
  \bibinfo {author} {\bibfnamefont {D.~M.}\ \bibnamefont {Eigler}},\ }\href
  {\doibase 10.1126/science.1101077} {\bibfield  {journal} {\bibinfo  {journal}
  {Science}\ }\textbf {\bibinfo {volume} {306}},\ \bibinfo {pages} {466}
  (\bibinfo {year} {2004})}\BibitemShut {NoStop}%
\bibitem [{\citenamefont {Hirjibehedin}\ \emph {et~al.}(2007)\citenamefont
  {Hirjibehedin}, \citenamefont {Lin}, \citenamefont {Otte}, \citenamefont
  {Ternes}, \citenamefont {Lutz}, \citenamefont {Jones},\ and\ \citenamefont
  {Heinrich}}]{Hirjibehedin2007}%
  \BibitemOpen
  \bibfield  {author} {\bibinfo {author} {\bibfnamefont {C.~F.}\ \bibnamefont
  {Hirjibehedin}}, \bibinfo {author} {\bibfnamefont {C.-Y.}\ \bibnamefont
  {Lin}}, \bibinfo {author} {\bibfnamefont {A.~F.}\ \bibnamefont {Otte}},
  \bibinfo {author} {\bibfnamefont {M.}~\bibnamefont {Ternes}}, \bibinfo
  {author} {\bibfnamefont {C.~P.}\ \bibnamefont {Lutz}}, \bibinfo {author}
  {\bibfnamefont {B.~A.}\ \bibnamefont {Jones}}, \ and\ \bibinfo {author}
  {\bibfnamefont {A.~J.}\ \bibnamefont {Heinrich}},\ }\href {\doibase
  10.1126/science.1146110} {\bibfield  {journal} {\bibinfo  {journal}
  {Science}\ }\textbf {\bibinfo {volume} {317}},\ \bibinfo {pages} {1199}
  (\bibinfo {year} {2007})}\BibitemShut {NoStop}%
\bibitem [{\citenamefont {Chen}\ \emph {et~al.}(2008)\citenamefont {Chen},
  \citenamefont {Fu}, \citenamefont {Ji}, \citenamefont {Zhang}, \citenamefont
  {Cheng}, \citenamefont {Ma}, \citenamefont {Zou}, \citenamefont {Duan},
  \citenamefont {Jia},\ and\ \citenamefont {Xue}}]{Chen2008}%
  \BibitemOpen
  \bibfield  {author} {\bibinfo {author} {\bibfnamefont {X.}~\bibnamefont
  {Chen}}, \bibinfo {author} {\bibfnamefont {Y.-S.}\ \bibnamefont {Fu}},
  \bibinfo {author} {\bibfnamefont {S.-H.}\ \bibnamefont {Ji}}, \bibinfo
  {author} {\bibfnamefont {T.}~\bibnamefont {Zhang}}, \bibinfo {author}
  {\bibfnamefont {P.}~\bibnamefont {Cheng}}, \bibinfo {author} {\bibfnamefont
  {X.-C.}\ \bibnamefont {Ma}}, \bibinfo {author} {\bibfnamefont {X.-L.}\
  \bibnamefont {Zou}}, \bibinfo {author} {\bibfnamefont {W.-H.}\ \bibnamefont
  {Duan}}, \bibinfo {author} {\bibfnamefont {J.-F.}\ \bibnamefont {Jia}}, \
  and\ \bibinfo {author} {\bibfnamefont {Q.-K.}\ \bibnamefont {Xue}},\ }\href
  {\doibase 10.1103/PhysRevLett.101.197208} {\bibfield  {journal} {\bibinfo
  {journal} {Phys. Rev. Lett.}\ }\textbf {\bibinfo {volume} {101}},\ \bibinfo
  {pages} {197208} (\bibinfo {year} {2008})}\BibitemShut {NoStop}%
\bibitem [{\citenamefont {Tsukahara}\ \emph {et~al.}(2009)\citenamefont
  {Tsukahara}, \citenamefont {Noto}, \citenamefont {Ohara}, \citenamefont
  {Shiraki}, \citenamefont {Takagi}, \citenamefont {Takata}, \citenamefont
  {Miyawaki}, \citenamefont {Taguchi}, \citenamefont {Chainani}, \citenamefont
  {Shin},\ and\ \citenamefont {Kawai}}]{Tsukahara2009}%
  \BibitemOpen
  \bibfield  {author} {\bibinfo {author} {\bibfnamefont {N.}~\bibnamefont
  {Tsukahara}}, \bibinfo {author} {\bibfnamefont {K.-i.}\ \bibnamefont {Noto}},
  \bibinfo {author} {\bibfnamefont {M.}~\bibnamefont {Ohara}}, \bibinfo
  {author} {\bibfnamefont {S.}~\bibnamefont {Shiraki}}, \bibinfo {author}
  {\bibfnamefont {N.}~\bibnamefont {Takagi}}, \bibinfo {author} {\bibfnamefont
  {Y.}~\bibnamefont {Takata}}, \bibinfo {author} {\bibfnamefont
  {J.}~\bibnamefont {Miyawaki}}, \bibinfo {author} {\bibfnamefont
  {M.}~\bibnamefont {Taguchi}}, \bibinfo {author} {\bibfnamefont
  {A.}~\bibnamefont {Chainani}}, \bibinfo {author} {\bibfnamefont
  {S.}~\bibnamefont {Shin}}, \ and\ \bibinfo {author} {\bibfnamefont
  {M.}~\bibnamefont {Kawai}},\ }\href {\doibase 10.1103/PhysRevLett.102.167203}
  {\bibfield  {journal} {\bibinfo  {journal} {Phys. Rev. Lett.}\ }\textbf
  {\bibinfo {volume} {102}},\ \bibinfo {pages} {167203} (\bibinfo {year}
  {2009})}\BibitemShut {NoStop}%
\bibitem [{\citenamefont {Khajetoorians}\ \emph {et~al.}(2015)\citenamefont
  {Khajetoorians}, \citenamefont {Valentyuk}, \citenamefont {Steinbrecher},
  \citenamefont {Schlenk}, \citenamefont {Shick}, \citenamefont {Kolorenc},
  \citenamefont {Lichtenstein}, \citenamefont {Wehling}, \citenamefont
  {Wiesendanger},\ and\ \citenamefont {Wiebe}}]{Khajetoorians2015}%
  \BibitemOpen
  \bibfield  {author} {\bibinfo {author} {\bibfnamefont {A.~A.}\ \bibnamefont
  {Khajetoorians}}, \bibinfo {author} {\bibfnamefont {M.}~\bibnamefont
  {Valentyuk}}, \bibinfo {author} {\bibfnamefont {M.}~\bibnamefont
  {Steinbrecher}}, \bibinfo {author} {\bibfnamefont {T.}~\bibnamefont
  {Schlenk}}, \bibinfo {author} {\bibfnamefont {A.}~\bibnamefont {Shick}},
  \bibinfo {author} {\bibfnamefont {J.}~\bibnamefont {Kolorenc}}, \bibinfo
  {author} {\bibfnamefont {A.~I.}\ \bibnamefont {Lichtenstein}}, \bibinfo
  {author} {\bibfnamefont {T.~O.}\ \bibnamefont {Wehling}}, \bibinfo {author}
  {\bibfnamefont {R.}~\bibnamefont {Wiesendanger}}, \ and\ \bibinfo {author}
  {\bibfnamefont {J.}~\bibnamefont {Wiebe}},\ }\href
  {http://dx.doi.org/10.1038/nnano.2015.193} {\bibfield  {journal} {\bibinfo
  {journal} {Nat. Nanotech.}\ }\textbf {\bibinfo {volume} {10}},\ \bibinfo
  {pages} {958} (\bibinfo {year} {2015})}\BibitemShut {NoStop}%
\bibitem [{\citenamefont {Jacobson}\ \emph {et~al.}(2015)\citenamefont
  {Jacobson}, \citenamefont {Herden}, \citenamefont {Muenks}, \citenamefont
  {Laskin}, \citenamefont {Brovko}, \citenamefont {Stepanyuk}, \citenamefont
  {Ternes},\ and\ \citenamefont {Kern}}]{Jacobson2015}%
  \BibitemOpen
  \bibfield  {author} {\bibinfo {author} {\bibfnamefont {P.}~\bibnamefont
  {Jacobson}}, \bibinfo {author} {\bibfnamefont {T.}~\bibnamefont {Herden}},
  \bibinfo {author} {\bibfnamefont {M.}~\bibnamefont {Muenks}}, \bibinfo
  {author} {\bibfnamefont {G.}~\bibnamefont {Laskin}}, \bibinfo {author}
  {\bibfnamefont {O.}~\bibnamefont {Brovko}}, \bibinfo {author} {\bibfnamefont
  {V.}~\bibnamefont {Stepanyuk}}, \bibinfo {author} {\bibfnamefont
  {M.}~\bibnamefont {Ternes}}, \ and\ \bibinfo {author} {\bibfnamefont
  {K.}~\bibnamefont {Kern}},\ }\href {http://dx.doi.org/10.1038/ncomms9536}
  {\bibfield  {journal} {\bibinfo  {journal} {Nat. Commun.}\ }\textbf {\bibinfo
  {volume} {6}},\ \bibinfo {pages} {8536} (\bibinfo {year} {2015})}\BibitemShut
  {NoStop}%
\bibitem [{\citenamefont {Loth}\ \emph
  {et~al.}(2010{\natexlab{a}})\citenamefont {Loth}, \citenamefont {von
  Bergmann}, \citenamefont {Ternes}, \citenamefont {Otte}, \citenamefont
  {Lutz},\ and\ \citenamefont {Heinrich}}]{Loth2010b}%
  \BibitemOpen
  \bibfield  {author} {\bibinfo {author} {\bibfnamefont {S.}~\bibnamefont
  {Loth}}, \bibinfo {author} {\bibfnamefont {K.}~\bibnamefont {von Bergmann}},
  \bibinfo {author} {\bibfnamefont {M.}~\bibnamefont {Ternes}}, \bibinfo
  {author} {\bibfnamefont {A.~F.}\ \bibnamefont {Otte}}, \bibinfo {author}
  {\bibfnamefont {C.~P.}\ \bibnamefont {Lutz}}, \ and\ \bibinfo {author}
  {\bibfnamefont {A.~J.}\ \bibnamefont {Heinrich}},\ }\href
  {http://dx.doi.org/10.1038/nphys1616} {\bibfield  {journal} {\bibinfo
  {journal} {Nat. Phys.}\ }\textbf {\bibinfo {volume} {6}},\ \bibinfo {pages}
  {340} (\bibinfo {year} {2010}{\natexlab{a}})}\BibitemShut {NoStop}%
\bibitem [{\citenamefont {Loth}\ \emph
  {et~al.}(2010{\natexlab{b}})\citenamefont {Loth}, \citenamefont {Etzkorn},
  \citenamefont {Lutz}, \citenamefont {Eigler},\ and\ \citenamefont
  {Heinrich}}]{Loth2010}%
  \BibitemOpen
  \bibfield  {author} {\bibinfo {author} {\bibfnamefont {S.}~\bibnamefont
  {Loth}}, \bibinfo {author} {\bibfnamefont {M.}~\bibnamefont {Etzkorn}},
  \bibinfo {author} {\bibfnamefont {C.~P.}\ \bibnamefont {Lutz}}, \bibinfo
  {author} {\bibfnamefont {D.~M.}\ \bibnamefont {Eigler}}, \ and\ \bibinfo
  {author} {\bibfnamefont {A.~J.}\ \bibnamefont {Heinrich}},\ }\href {\doibase
  10.1126/science.1191688} {\bibfield  {journal} {\bibinfo  {journal}
  {Science}\ }\textbf {\bibinfo {volume} {329}},\ \bibinfo {pages} {1628}
  (\bibinfo {year} {2010}{\natexlab{b}})}\BibitemShut {NoStop}%
\bibitem [{\citenamefont {Yan}\ \emph {et~al.}(2015)\citenamefont {Yan},
  \citenamefont {Choi}, \citenamefont {J.}, \citenamefont {Rolf-Pissarczyk},\
  and\ \citenamefont {Loth}}]{Yan2015}%
  \BibitemOpen
  \bibfield  {author} {\bibinfo {author} {\bibfnamefont {S.}~\bibnamefont
  {Yan}}, \bibinfo {author} {\bibfnamefont {D.-J.}\ \bibnamefont {Choi}},
  \bibinfo {author} {\bibfnamefont {B.~A.}\ \bibnamefont {J.}}, \bibinfo
  {author} {\bibfnamefont {S.}~\bibnamefont {Rolf-Pissarczyk}}, \ and\ \bibinfo
  {author} {\bibfnamefont {S.}~\bibnamefont {Loth}},\ }\href
  {http://dx.doi.org/10.1038/nnano.2014.281} {\bibfield  {journal} {\bibinfo
  {journal} {Nat. Nanotech.}\ }\textbf {\bibinfo {volume} {10}},\ \bibinfo
  {pages} {40} (\bibinfo {year} {2015})}\BibitemShut {NoStop}%
\bibitem [{\citenamefont {Heinrich}\ \emph {et~al.}(2013)\citenamefont
  {Heinrich}, \citenamefont {Braun}, \citenamefont {Pascual},\ and\
  \citenamefont {Franke}}]{Heinrich2013}%
  \BibitemOpen
  \bibfield  {author} {\bibinfo {author} {\bibfnamefont {B.~W.}\ \bibnamefont
  {Heinrich}}, \bibinfo {author} {\bibfnamefont {L.}~\bibnamefont {Braun}},
  \bibinfo {author} {\bibfnamefont {J.~I.}\ \bibnamefont {Pascual}}, \ and\
  \bibinfo {author} {\bibfnamefont {K.~J.}\ \bibnamefont {Franke}},\ }\href
  {\doibase 10.1038/nphys2794} {\bibfield  {journal} {\bibinfo  {journal} {Nat.
  Phys.}\ }\textbf {\bibinfo {volume} {9}},\ \bibinfo {pages} {765 } (\bibinfo
  {year} {2013})}\BibitemShut {NoStop}%
\bibitem [{\citenamefont {Wagner}\ and\ \citenamefont
  {Temirov}(2015)}]{Wagner2015}%
  \BibitemOpen
  \bibfield  {author} {\bibinfo {author} {\bibfnamefont {C.}~\bibnamefont
  {Wagner}}\ and\ \bibinfo {author} {\bibfnamefont {R.}~\bibnamefont
  {Temirov}},\ }\href {\doibase
  http://dx.doi.org/10.1016/j.progsurf.2015.01.001} {\bibfield  {journal}
  {\bibinfo  {journal} {Progress in Surface Science}\ }\textbf {\bibinfo
  {volume} {90}},\ \bibinfo {pages} {194 } (\bibinfo {year}
  {2015})}\BibitemShut {NoStop}%
\bibitem [{\citenamefont {Temirov}\ \emph {et~al.}(2008)\citenamefont
  {Temirov}, \citenamefont {Soubatch}, \citenamefont {Neucheva}, \citenamefont
  {Lassise},\ and\ \citenamefont {Tautz}}]{Temirov2008}%
  \BibitemOpen
  \bibfield  {author} {\bibinfo {author} {\bibfnamefont {R.}~\bibnamefont
  {Temirov}}, \bibinfo {author} {\bibfnamefont {S.}~\bibnamefont {Soubatch}},
  \bibinfo {author} {\bibfnamefont {O.}~\bibnamefont {Neucheva}}, \bibinfo
  {author} {\bibfnamefont {A.~C.}\ \bibnamefont {Lassise}}, \ and\ \bibinfo
  {author} {\bibfnamefont {F.~S.}\ \bibnamefont {Tautz}},\ }\href
  {http://stacks.iop.org/1367-2630/10/i=5/a=053012} {\bibfield  {journal}
  {\bibinfo  {journal} {New Journal of Physics}\ }\textbf {\bibinfo {volume}
  {10}},\ \bibinfo {pages} {053012} (\bibinfo {year} {2008})}\BibitemShut
  {NoStop}%
\bibitem [{\citenamefont {Weiss}\ \emph {et~al.}(2010)\citenamefont {Weiss},
  \citenamefont {Wagner}, \citenamefont {Kleimann}, \citenamefont {Rohlfing},
  \citenamefont {Tautz},\ and\ \citenamefont {Temirov}}]{Weiss2010}%
  \BibitemOpen
  \bibfield  {author} {\bibinfo {author} {\bibfnamefont {C.}~\bibnamefont
  {Weiss}}, \bibinfo {author} {\bibfnamefont {C.}~\bibnamefont {Wagner}},
  \bibinfo {author} {\bibfnamefont {C.}~\bibnamefont {Kleimann}}, \bibinfo
  {author} {\bibfnamefont {M.}~\bibnamefont {Rohlfing}}, \bibinfo {author}
  {\bibfnamefont {F.~S.}\ \bibnamefont {Tautz}}, \ and\ \bibinfo {author}
  {\bibfnamefont {R.}~\bibnamefont {Temirov}},\ }\href {\doibase
  10.1103/PhysRevLett.105.086103} {\bibfield  {journal} {\bibinfo  {journal}
  {Phys. Rev. Lett.}\ }\textbf {\bibinfo {volume} {105}},\ \bibinfo {pages}
  {086103} (\bibinfo {year} {2010})}\BibitemShut {NoStop}%
\bibitem [{\citenamefont {Gross}\ \emph {et~al.}(2011)\citenamefont {Gross},
  \citenamefont {Moll}, \citenamefont {Mohn}, \citenamefont {Curioni},
  \citenamefont {Meyer}, \citenamefont {Hanke},\ and\ \citenamefont
  {Persson}}]{Gross2011}%
  \BibitemOpen
  \bibfield  {author} {\bibinfo {author} {\bibfnamefont {L.}~\bibnamefont
  {Gross}}, \bibinfo {author} {\bibfnamefont {N.}~\bibnamefont {Moll}},
  \bibinfo {author} {\bibfnamefont {F.}~\bibnamefont {Mohn}}, \bibinfo {author}
  {\bibfnamefont {A.}~\bibnamefont {Curioni}}, \bibinfo {author} {\bibfnamefont
  {G.}~\bibnamefont {Meyer}}, \bibinfo {author} {\bibfnamefont
  {F.}~\bibnamefont {Hanke}}, \ and\ \bibinfo {author} {\bibfnamefont
  {M.}~\bibnamefont {Persson}},\ }\href {\doibase
  10.1103/PhysRevLett.107.086101} {\bibfield  {journal} {\bibinfo  {journal}
  {Phys. Rev. Lett.}\ }\textbf {\bibinfo {volume} {107}},\ \bibinfo {pages}
  {086101} (\bibinfo {year} {2011})}\BibitemShut {NoStop}%
\bibitem [{\citenamefont {Chiang}\ \emph {et~al.}(2014)\citenamefont {Chiang},
  \citenamefont {Xu}, \citenamefont {Han},\ and\ \citenamefont {Ho}}]{Ho2014}%
  \BibitemOpen
  \bibfield  {author} {\bibinfo {author} {\bibfnamefont {C.-l.}\ \bibnamefont
  {Chiang}}, \bibinfo {author} {\bibfnamefont {C.}~\bibnamefont {Xu}}, \bibinfo
  {author} {\bibfnamefont {Z.}~\bibnamefont {Han}}, \ and\ \bibinfo {author}
  {\bibfnamefont {W.}~\bibnamefont {Ho}},\ }\href {\doibase
  10.1126/science.1253405} {\bibfield  {journal} {\bibinfo  {journal}
  {Science}\ }\textbf {\bibinfo {volume} {344}},\ \bibinfo {pages} {885}
  (\bibinfo {year} {2014})}\BibitemShut {NoStop}%
\bibitem [{\citenamefont {Kichin}\ \emph {et~al.}()\citenamefont {Kichin},
  \citenamefont {Weiss}, \citenamefont {Wagner}, \citenamefont {Tautz},\ and\
  \citenamefont {Temirov}}]{Kichin2011}%
  \BibitemOpen
  \bibfield  {author} {\bibinfo {author} {\bibfnamefont {G.}~\bibnamefont
  {Kichin}}, \bibinfo {author} {\bibfnamefont {C.}~\bibnamefont {Weiss}},
  \bibinfo {author} {\bibfnamefont {C.}~\bibnamefont {Wagner}}, \bibinfo
  {author} {\bibfnamefont {F.~S.}\ \bibnamefont {Tautz}}, \ and\ \bibinfo
  {author} {\bibfnamefont {R.}~\bibnamefont {Temirov}},\ }\href@noop {} {\
  }\BibitemShut {NoStop}%
\bibitem [{\citenamefont {Kichin}\ \emph {et~al.}(2013)\citenamefont {Kichin},
  \citenamefont {Wagner}, \citenamefont {Tautz},\ and\ \citenamefont
  {Temirov}}]{Kichin2013}%
  \BibitemOpen
  \bibfield  {author} {\bibinfo {author} {\bibfnamefont {G.}~\bibnamefont
  {Kichin}}, \bibinfo {author} {\bibfnamefont {C.}~\bibnamefont {Wagner}},
  \bibinfo {author} {\bibfnamefont {F.~S.}\ \bibnamefont {Tautz}}, \ and\
  \bibinfo {author} {\bibfnamefont {R.}~\bibnamefont {Temirov}},\ }\href
  {\doibase 10.1103/PhysRevB.87.081408} {\bibfield  {journal} {\bibinfo
  {journal} {Phys. Rev. B}\ }\textbf {\bibinfo {volume} {87}},\ \bibinfo
  {pages} {081408} (\bibinfo {year} {2013})}\BibitemShut {NoStop}%
\bibitem [{\citenamefont {Hapala}\ \emph {et~al.}(2014)\citenamefont {Hapala},
  \citenamefont {Kichin}, \citenamefont {Wagner}, \citenamefont {Tautz},
  \citenamefont {Temirov},\ and\ \citenamefont {Jel\'{\i}nek}}]{Hapala2014}%
  \BibitemOpen
  \bibfield  {author} {\bibinfo {author} {\bibfnamefont {P.}~\bibnamefont
  {Hapala}}, \bibinfo {author} {\bibfnamefont {G.}~\bibnamefont {Kichin}},
  \bibinfo {author} {\bibfnamefont {C.}~\bibnamefont {Wagner}}, \bibinfo
  {author} {\bibfnamefont {F.~S.}\ \bibnamefont {Tautz}}, \bibinfo {author}
  {\bibfnamefont {R.}~\bibnamefont {Temirov}}, \ and\ \bibinfo {author}
  {\bibfnamefont {P.}~\bibnamefont {Jel\'{\i}nek}},\ }\href {\doibase
  10.1103/PhysRevB.90.085421} {\bibfield  {journal} {\bibinfo  {journal} {Phys.
  Rev. B}\ }\textbf {\bibinfo {volume} {90}},\ \bibinfo {pages} {085421}
  (\bibinfo {year} {2014})}\BibitemShut {NoStop}%
\bibitem [{\citenamefont {Gross}\ \emph {et~al.}(2009)\citenamefont {Gross},
  \citenamefont {Mohn}, \citenamefont {Moll}, \citenamefont {Liljeroth},\ and\
  \citenamefont {Meyer}}]{Gross2009}%
  \BibitemOpen
  \bibfield  {author} {\bibinfo {author} {\bibfnamefont {L.}~\bibnamefont
  {Gross}}, \bibinfo {author} {\bibfnamefont {F.}~\bibnamefont {Mohn}},
  \bibinfo {author} {\bibfnamefont {N.}~\bibnamefont {Moll}}, \bibinfo {author}
  {\bibfnamefont {P.}~\bibnamefont {Liljeroth}}, \ and\ \bibinfo {author}
  {\bibfnamefont {G.}~\bibnamefont {Meyer}},\ }\href {\doibase
  10.1126/science.1176210} {\bibfield  {journal} {\bibinfo  {journal}
  {Science}\ }\textbf {\bibinfo {volume} {325}},\ \bibinfo {pages} {1110}
  (\bibinfo {year} {2009})}\BibitemShut {NoStop}%
\bibitem [{\citenamefont {Guo}\ \emph {et~al.}(2016)\citenamefont {Guo},
  \citenamefont {L{\"u}}, \citenamefont {Feng}, \citenamefont {Chen},
  \citenamefont {Peng}, \citenamefont {Lin}, \citenamefont {Meng},
  \citenamefont {Wang}, \citenamefont {Li}, \citenamefont {Wang},\ and\
  \citenamefont {Jiang}}]{Guo2016}%
  \BibitemOpen
  \bibfield  {author} {\bibinfo {author} {\bibfnamefont {J.}~\bibnamefont
  {Guo}}, \bibinfo {author} {\bibfnamefont {J.-T.}\ \bibnamefont {L{\"u}}},
  \bibinfo {author} {\bibfnamefont {Y.}~\bibnamefont {Feng}}, \bibinfo {author}
  {\bibfnamefont {J.}~\bibnamefont {Chen}}, \bibinfo {author} {\bibfnamefont
  {J.}~\bibnamefont {Peng}}, \bibinfo {author} {\bibfnamefont {Z.}~\bibnamefont
  {Lin}}, \bibinfo {author} {\bibfnamefont {X.}~\bibnamefont {Meng}}, \bibinfo
  {author} {\bibfnamefont {Z.}~\bibnamefont {Wang}}, \bibinfo {author}
  {\bibfnamefont {X.-Z.}\ \bibnamefont {Li}}, \bibinfo {author} {\bibfnamefont
  {E.-G.}\ \bibnamefont {Wang}}, \ and\ \bibinfo {author} {\bibfnamefont
  {Y.}~\bibnamefont {Jiang}},\ }\href {\doibase 10.1126/science.aaf2042}
  {\bibfield  {journal} {\bibinfo  {journal} {Science}\ }\textbf {\bibinfo
  {volume} {352}},\ \bibinfo {pages} {321} (\bibinfo {year}
  {2016})}\BibitemShut {NoStop}%
\bibitem [{\citenamefont {Wagner}\ \emph {et~al.}(2015)\citenamefont {Wagner},
  \citenamefont {Green}, \citenamefont {Leinen}, \citenamefont {Deilmann},
  \citenamefont {Kr\"uger}, \citenamefont {Rohlfing}, \citenamefont {Temirov},\
  and\ \citenamefont {Tautz}}]{Wagner2015b}%
  \BibitemOpen
  \bibfield  {author} {\bibinfo {author} {\bibfnamefont {C.}~\bibnamefont
  {Wagner}}, \bibinfo {author} {\bibfnamefont {M.~F.~B.}\ \bibnamefont
  {Green}}, \bibinfo {author} {\bibfnamefont {P.}~\bibnamefont {Leinen}},
  \bibinfo {author} {\bibfnamefont {T.}~\bibnamefont {Deilmann}}, \bibinfo
  {author} {\bibfnamefont {P.}~\bibnamefont {Kr\"uger}}, \bibinfo {author}
  {\bibfnamefont {M.}~\bibnamefont {Rohlfing}}, \bibinfo {author}
  {\bibfnamefont {R.}~\bibnamefont {Temirov}}, \ and\ \bibinfo {author}
  {\bibfnamefont {F.~S.}\ \bibnamefont {Tautz}},\ }\href {\doibase
  10.1103/PhysRevLett.115.026101} {\bibfield  {journal} {\bibinfo  {journal}
  {Phys. Rev. Lett.}\ }\textbf {\bibinfo {volume} {115}},\ \bibinfo {pages}
  {026101} (\bibinfo {year} {2015})}\BibitemShut {NoStop}%
\bibitem [{\citenamefont {Zhen-Feng}\ \emph {et~al.}(2003)\citenamefont
  {Zhen-Feng}, \citenamefont {Yaoming}, \citenamefont {Wen-Lin},\ and\
  \citenamefont {Henry~F.}}]{Feng2003}%
  \BibitemOpen
  \bibfield  {author} {\bibinfo {author} {\bibfnamefont {X.}~\bibnamefont
  {Zhen-Feng}}, \bibinfo {author} {\bibfnamefont {X.}~\bibnamefont {Yaoming}},
  \bibinfo {author} {\bibfnamefont {F.}~\bibnamefont {Wen-Lin}}, \ and\
  \bibinfo {author} {\bibfnamefont {S.}~\bibnamefont {Henry~F.}},\ }\href
  {\doibase 10.1021/jp0219855} {\bibfield  {journal} {\bibinfo  {journal} {J.
  Phys. Chem. A}\ }\textbf {\bibinfo {volume} {107}},\ \bibinfo {pages} {2716}
  (\bibinfo {year} {2003})}\BibitemShut {NoStop}%
\bibitem [{SI()}]{SI}%
  \BibitemOpen
  \href@noop {} {}\bibinfo {howpublished} {Materials and methods are available
  as supplementary materials}\BibitemShut {NoStop}%
\bibitem [{\citenamefont {Bachellier}\ \emph {et~al.}(2016)\citenamefont
  {Bachellier}, \citenamefont {Ormaza}, \citenamefont {Faraggi}, \citenamefont
  {Verlhac}, \citenamefont {V\'erot}, \citenamefont {Le~Bahers}, \citenamefont
  {Bocquet},\ and\ \citenamefont {Limot}}]{Bachellier2016}%
  \BibitemOpen
  \bibfield  {author} {\bibinfo {author} {\bibfnamefont {N.}~\bibnamefont
  {Bachellier}}, \bibinfo {author} {\bibfnamefont {M.}~\bibnamefont {Ormaza}},
  \bibinfo {author} {\bibfnamefont {M.}~\bibnamefont {Faraggi}}, \bibinfo
  {author} {\bibfnamefont {B.}~\bibnamefont {Verlhac}}, \bibinfo {author}
  {\bibfnamefont {M.}~\bibnamefont {V\'erot}}, \bibinfo {author} {\bibfnamefont
  {T.}~\bibnamefont {Le~Bahers}}, \bibinfo {author} {\bibfnamefont {M.-L.}\
  \bibnamefont {Bocquet}}, \ and\ \bibinfo {author} {\bibfnamefont
  {L.}~\bibnamefont {Limot}},\ }\href {\doibase 10.1103/PhysRevB.93.195403}
  {\bibfield  {journal} {\bibinfo  {journal} {Phys. Rev. B}\ }\textbf {\bibinfo
  {volume} {93}},\ \bibinfo {pages} {195403} (\bibinfo {year}
  {2016})}\BibitemShut {NoStop}%
\bibitem [{\citenamefont {Pugmire}\ \emph {et~al.}(2001)\citenamefont
  {Pugmire}, \citenamefont {Woodbridge}, \citenamefont {Boag},\ and\
  \citenamefont {Langell}}]{Pugmire2001}%
  \BibitemOpen
  \bibfield  {author} {\bibinfo {author} {\bibfnamefont {D.}~\bibnamefont
  {Pugmire}}, \bibinfo {author} {\bibfnamefont {C.}~\bibnamefont {Woodbridge}},
  \bibinfo {author} {\bibfnamefont {N.}~\bibnamefont {Boag}}, \ and\ \bibinfo
  {author} {\bibfnamefont {M.}~\bibnamefont {Langell}},\ }\href {\doibase
  http://dx.doi.org/10.1016/S0039-6028(00)00939-0} {\bibfield  {journal}
  {\bibinfo  {journal} {Surf. Sci.}\ }\textbf {\bibinfo {volume} {472}},\
  \bibinfo {pages} {155 } (\bibinfo {year} {2001})}\BibitemShut {NoStop}%
\bibitem [{\citenamefont {Lorente}\ and\ \citenamefont
  {Gauyacq}(2009)}]{Lorente2009}%
  \BibitemOpen
  \bibfield  {author} {\bibinfo {author} {\bibfnamefont {N.}~\bibnamefont
  {Lorente}}\ and\ \bibinfo {author} {\bibfnamefont {J.-P.}\ \bibnamefont
  {Gauyacq}},\ }\href {\doibase 10.1103/PhysRevLett.103.176601} {\bibfield
  {journal} {\bibinfo  {journal} {Phys. Rev. Lett.}\ }\textbf {\bibinfo
  {volume} {103}},\ \bibinfo {pages} {176601} (\bibinfo {year}
  {2009})}\BibitemShut {NoStop}%
\bibitem [{\citenamefont {Gauyacq}\ \emph {et~al.}(2012)\citenamefont
  {Gauyacq}, \citenamefont {Lorente},\ and\ \citenamefont
  {Novaes}}]{Gauyacq2012}%
  \BibitemOpen
  \bibfield  {author} {\bibinfo {author} {\bibfnamefont {J.-P.}\ \bibnamefont
  {Gauyacq}}, \bibinfo {author} {\bibfnamefont {N.}~\bibnamefont {Lorente}}, \
  and\ \bibinfo {author} {\bibfnamefont {F.~D.}\ \bibnamefont {Novaes}},\
  }\href {\doibase http://dx.doi.org/10.1016/j.progsurf.2012.05.003} {\bibfield
   {journal} {\bibinfo  {journal} {Progress in Surface Science}\ }\textbf
  {\bibinfo {volume} {87}},\ \bibinfo {pages} {63 } (\bibinfo {year}
  {2012})}\BibitemShut {NoStop}%
\bibitem [{\citenamefont {Miyamachi}\ \emph {et~al.}(2013)\citenamefont
  {Miyamachi}, \citenamefont {Schuh}, \citenamefont {Markl}, \citenamefont
  {Bresch}, \citenamefont {Balashov}, \citenamefont {Stohr}, \citenamefont
  {Karlewski}, \citenamefont {Andre}, \citenamefont {Marthaler}, \citenamefont
  {Hoffmann}, \citenamefont {Geilhufe}, \citenamefont {Ostanin}, \citenamefont
  {Hergert}, \citenamefont {Mertig}, \citenamefont {Schon}, \citenamefont
  {Ernst},\ and\ \citenamefont {Wulfhekel}}]{Miyamachi2013}%
  \BibitemOpen
  \bibfield  {author} {\bibinfo {author} {\bibfnamefont {T.}~\bibnamefont
  {Miyamachi}}, \bibinfo {author} {\bibfnamefont {T.}~\bibnamefont {Schuh}},
  \bibinfo {author} {\bibfnamefont {T.}~\bibnamefont {Markl}}, \bibinfo
  {author} {\bibfnamefont {C.}~\bibnamefont {Bresch}}, \bibinfo {author}
  {\bibfnamefont {T.}~\bibnamefont {Balashov}}, \bibinfo {author}
  {\bibfnamefont {A.}~\bibnamefont {Stohr}}, \bibinfo {author} {\bibfnamefont
  {C.}~\bibnamefont {Karlewski}}, \bibinfo {author} {\bibfnamefont
  {S.}~\bibnamefont {Andre}}, \bibinfo {author} {\bibfnamefont
  {M.}~\bibnamefont {Marthaler}}, \bibinfo {author} {\bibfnamefont
  {M.}~\bibnamefont {Hoffmann}}, \bibinfo {author} {\bibfnamefont
  {M.}~\bibnamefont {Geilhufe}}, \bibinfo {author} {\bibfnamefont
  {S.}~\bibnamefont {Ostanin}}, \bibinfo {author} {\bibfnamefont
  {W.}~\bibnamefont {Hergert}}, \bibinfo {author} {\bibfnamefont
  {I.}~\bibnamefont {Mertig}}, \bibinfo {author} {\bibfnamefont
  {G.}~\bibnamefont {Schon}}, \bibinfo {author} {\bibfnamefont
  {A.}~\bibnamefont {Ernst}}, \ and\ \bibinfo {author} {\bibfnamefont
  {W.}~\bibnamefont {Wulfhekel}},\ }\href@noop {} {\bibfield  {journal}
  {\bibinfo  {journal} {Nature}\ }\textbf {\bibinfo {volume} {503}},\ \bibinfo
  {pages} {242} (\bibinfo {year} {2013})}\BibitemShut {NoStop}%
\bibitem [{\citenamefont {Ormaza}\ \emph {et~al.}(2015)\citenamefont {Ormaza},
  \citenamefont {Abufager}, \citenamefont {Bachellier}, \citenamefont {Robles},
  \citenamefont {Verot}, \citenamefont {Le~Bahers}, \citenamefont {Bocquet},
  \citenamefont {Lorente},\ and\ \citenamefont {Limot}}]{Ormaza2015}%
  \BibitemOpen
  \bibfield  {author} {\bibinfo {author} {\bibfnamefont {M.}~\bibnamefont
  {Ormaza}}, \bibinfo {author} {\bibfnamefont {P.}~\bibnamefont {Abufager}},
  \bibinfo {author} {\bibfnamefont {N.}~\bibnamefont {Bachellier}}, \bibinfo
  {author} {\bibfnamefont {R.}~\bibnamefont {Robles}}, \bibinfo {author}
  {\bibfnamefont {M.}~\bibnamefont {Verot}}, \bibinfo {author} {\bibfnamefont
  {T.}~\bibnamefont {Le~Bahers}}, \bibinfo {author} {\bibfnamefont {M.-L.}\
  \bibnamefont {Bocquet}}, \bibinfo {author} {\bibfnamefont {N.}~\bibnamefont
  {Lorente}}, \ and\ \bibinfo {author} {\bibfnamefont {L.}~\bibnamefont
  {Limot}},\ }\href@noop {} {\bibfield  {journal} {\bibinfo  {journal} {J.
  Phys. Chem. Lett.}\ }\textbf {\bibinfo {volume} {6}},\ \bibinfo {pages} {395}
  (\bibinfo {year} {2015})}\BibitemShut {NoStop}%
\bibitem [{\citenamefont {Prins}\ \emph {et~al.}(1967)\citenamefont {Prins},
  \citenamefont {van Voorst},\ and\ \citenamefont {Schinkel}}]{Prins1967}%
  \BibitemOpen
  \bibfield  {author} {\bibinfo {author} {\bibfnamefont {R.}~\bibnamefont
  {Prins}}, \bibinfo {author} {\bibfnamefont {J.}~\bibnamefont {van Voorst}}, \
  and\ \bibinfo {author} {\bibfnamefont {C.}~\bibnamefont {Schinkel}},\ }\href
  {\doibase http://dx.doi.org/10.1016/0009-2614(67)80067-8} {\bibfield
  {journal} {\bibinfo  {journal} {Chem. Phys. Lett.}\ }\textbf {\bibinfo
  {volume} {1}},\ \bibinfo {pages} {54 } (\bibinfo {year} {1967})}\BibitemShut
  {NoStop}%
\bibitem [{\citenamefont {Baltzer}\ \emph {et~al.}(1988)\citenamefont
  {Baltzer}, \citenamefont {Furrer}, \citenamefont {Hulliger},\ and\
  \citenamefont {Stebler}}]{Baltzer1988}%
  \BibitemOpen
  \bibfield  {author} {\bibinfo {author} {\bibfnamefont {P.}~\bibnamefont
  {Baltzer}}, \bibinfo {author} {\bibfnamefont {A.}~\bibnamefont {Furrer}},
  \bibinfo {author} {\bibfnamefont {J.}~\bibnamefont {Hulliger}}, \ and\
  \bibinfo {author} {\bibfnamefont {A.}~\bibnamefont {Stebler}},\ }\href
  {\doibase 10.1021/ic00282a007} {\bibfield  {journal} {\bibinfo  {journal}
  {Inorg. Chem.}\ }\textbf {\bibinfo {volume} {27}},\ \bibinfo {pages} {1543}
  (\bibinfo {year} {1988})}\BibitemShut {NoStop}%
\bibitem [{\citenamefont {Li}\ \emph {et~al.}(1992)\citenamefont {Li},
  \citenamefont {Hamrick}, \citenamefont {Zee},\ and\ \citenamefont
  {Jr.}}]{Li1992}%
  \BibitemOpen
  \bibfield  {author} {\bibinfo {author} {\bibfnamefont {S.}~\bibnamefont
  {Li}}, \bibinfo {author} {\bibfnamefont {Y.~M.}\ \bibnamefont {Hamrick}},
  \bibinfo {author} {\bibfnamefont {R.~J.~V.}\ \bibnamefont {Zee}}, \ and\
  \bibinfo {author} {\bibfnamefont {W.~W.}\ \bibnamefont {Jr.}},\ }\href
  {\doibase 10.1021/ja00037a078} {\bibfield  {journal} {\bibinfo  {journal} {J.
  Am. Chem. Soc.}\ }\textbf {\bibinfo {volume} {114}},\ \bibinfo {pages} {4433}
  (\bibinfo {year} {1992})}\BibitemShut {NoStop}%
\bibitem [{\citenamefont {Vaara}\ \emph {et~al.}(2015)\citenamefont {Vaara},
  \citenamefont {Rouf},\ and\ \citenamefont {MareÅ¡}}]{Vaara2015}%
  \BibitemOpen
  \bibfield  {author} {\bibinfo {author} {\bibfnamefont {J.}~\bibnamefont
  {Vaara}}, \bibinfo {author} {\bibfnamefont {S.~A.}\ \bibnamefont {Rouf}}, \
  and\ \bibinfo {author} {\bibfnamefont {J.}~\bibnamefont {MareÅ¡}},\ }\href
  {\doibase 10.1021/acs.jctc.5b00656} {\bibfield  {journal} {\bibinfo
  {journal} {J. Chem. Theory Comput.}\ }\textbf {\bibinfo {volume} {11}},\
  \bibinfo {pages} {4840} (\bibinfo {year} {2015})}\BibitemShut {NoStop}%
\bibitem [{\citenamefont {Eigler}\ \emph {et~al.}(1991)\citenamefont {Eigler},
  \citenamefont {Lutz},\ and\ \citenamefont {Rudge}}]{Eigler1991}%
  \BibitemOpen
  \bibfield  {author} {\bibinfo {author} {\bibfnamefont {D.}~\bibnamefont
  {Eigler}}, \bibinfo {author} {\bibfnamefont {C.}~\bibnamefont {Lutz}}, \ and\
  \bibinfo {author} {\bibfnamefont {W.}~\bibnamefont {Rudge}},\ }\href
  {\doibase http://dx.doi.org/10.1038/352600a0} {\bibfield  {journal} {\bibinfo
   {journal} {Nature}\ }\textbf {\bibinfo {volume} {352}},\ \bibinfo {pages}
  {600 } (\bibinfo {year} {1991})}\BibitemShut {NoStop}%
\bibitem [{\citenamefont {Bartels}\ \emph {et~al.}(1998)\citenamefont
  {Bartels}, \citenamefont {Meyer}, \citenamefont {Rieder}, \citenamefont
  {Velic}, \citenamefont {Knoesel}, \citenamefont {Hotzel}, \citenamefont
  {Wolf},\ and\ \citenamefont {Ertl}}]{Bartels1998}%
  \BibitemOpen
  \bibfield  {author} {\bibinfo {author} {\bibfnamefont {L.}~\bibnamefont
  {Bartels}}, \bibinfo {author} {\bibfnamefont {G.}~\bibnamefont {Meyer}},
  \bibinfo {author} {\bibfnamefont {K.-H.}\ \bibnamefont {Rieder}}, \bibinfo
  {author} {\bibfnamefont {D.}~\bibnamefont {Velic}}, \bibinfo {author}
  {\bibfnamefont {E.}~\bibnamefont {Knoesel}}, \bibinfo {author} {\bibfnamefont
  {A.}~\bibnamefont {Hotzel}}, \bibinfo {author} {\bibfnamefont
  {M.}~\bibnamefont {Wolf}}, \ and\ \bibinfo {author} {\bibfnamefont
  {G.}~\bibnamefont {Ertl}},\ }\href {\doibase 10.1103/PhysRevLett.80.2004}
  {\bibfield  {journal} {\bibinfo  {journal} {Phys. Rev. Lett.}\ }\textbf
  {\bibinfo {volume} {80}},\ \bibinfo {pages} {2004} (\bibinfo {year}
  {1998})}\BibitemShut {NoStop}%
\bibitem [{\citenamefont {Ormaza}\ \emph {et~al.}(2016)\citenamefont {Ormaza},
  \citenamefont {Robles}, \citenamefont {Bachellier}, \citenamefont {Abufager},
  \citenamefont {Lorente},\ and\ \citenamefont {Limot}}]{Ormaza2016}%
  \BibitemOpen
  \bibfield  {author} {\bibinfo {author} {\bibfnamefont {M.}~\bibnamefont
  {Ormaza}}, \bibinfo {author} {\bibfnamefont {R.}~\bibnamefont {Robles}},
  \bibinfo {author} {\bibfnamefont {N.}~\bibnamefont {Bachellier}}, \bibinfo
  {author} {\bibfnamefont {P.}~\bibnamefont {Abufager}}, \bibinfo {author}
  {\bibfnamefont {N.}~\bibnamefont {Lorente}}, \ and\ \bibinfo {author}
  {\bibfnamefont {L.}~\bibnamefont {Limot}},\ }\href {\doibase
  10.1021/acs.nanolett.5b04280} {\bibfield  {journal} {\bibinfo  {journal}
  {Nano Lett.}\ }\textbf {\bibinfo {volume} {16}},\ \bibinfo {pages} {588}
  (\bibinfo {year} {2016})}\BibitemShut {NoStop}%
\bibitem [{\citenamefont {Heinrich}\ \emph {et~al.}(2011)\citenamefont
  {Heinrich}, \citenamefont {Rastei}, \citenamefont {Choi}, \citenamefont
  {Frederiksen},\ and\ \citenamefont {Limot}}]{Heinrich2011}%
  \BibitemOpen
  \bibfield  {author} {\bibinfo {author} {\bibfnamefont {B.~W.}\ \bibnamefont
  {Heinrich}}, \bibinfo {author} {\bibfnamefont {M.~V.}\ \bibnamefont
  {Rastei}}, \bibinfo {author} {\bibfnamefont {D.-J.}\ \bibnamefont {Choi}},
  \bibinfo {author} {\bibfnamefont {T.}~\bibnamefont {Frederiksen}}, \ and\
  \bibinfo {author} {\bibfnamefont {L.}~\bibnamefont {Limot}},\ }\href
  {\doibase 10.1103/PhysRevLett.107.246801} {\bibfield  {journal} {\bibinfo
  {journal} {Phys. Rev. Lett.}\ }\textbf {\bibinfo {volume} {107}},\ \bibinfo
  {pages} {246801} (\bibinfo {year} {2011})}\BibitemShut {NoStop}%
\bibitem [{\citenamefont {Heinrich}\ \emph {et~al.}(2015)\citenamefont
  {Heinrich}, \citenamefont {Braun}, \citenamefont {Pascual},\ and\
  \citenamefont {Franke}}]{Heinrich2015}%
  \BibitemOpen
  \bibfield  {author} {\bibinfo {author} {\bibfnamefont {B.~W.}\ \bibnamefont
  {Heinrich}}, \bibinfo {author} {\bibfnamefont {L.}~\bibnamefont {Braun}},
  \bibinfo {author} {\bibfnamefont {J.~I.}\ \bibnamefont {Pascual}}, \ and\
  \bibinfo {author} {\bibfnamefont {K.~J.}\ \bibnamefont {Franke}},\ }\href
  {\doibase 10.1021/acs.nanolett.5b00987} {\bibfield  {journal} {\bibinfo
  {journal} {Nano Lett.}\ }\textbf {\bibinfo {volume} {15}},\ \bibinfo {pages}
  {4024} (\bibinfo {year} {2015})}\BibitemShut {NoStop}%
\bibitem [{\citenamefont {Ohresser}\ \emph {et~al.}(2014)\citenamefont
  {Ohresser}, \citenamefont {Otero}, \citenamefont {Choueikani}, \citenamefont
  {Chen}, \citenamefont {Stanescu}, \citenamefont {Deschamps}, \citenamefont
  {Moreno}, \citenamefont {Polack}, \citenamefont {Lagarde}, \citenamefont
  {Daguerre}, \citenamefont {Marteau}, \citenamefont {Scheurer}, \citenamefont
  {Joly}, \citenamefont {Kappler}, \citenamefont {Muller}, \citenamefont
  {Bunau},\ and\ \citenamefont {Sainctavit}}]{deimos}%
  \BibitemOpen
  \bibfield  {author} {\bibinfo {author} {\bibfnamefont {P.}~\bibnamefont
  {Ohresser}}, \bibinfo {author} {\bibfnamefont {E.}~\bibnamefont {Otero}},
  \bibinfo {author} {\bibfnamefont {F.}~\bibnamefont {Choueikani}}, \bibinfo
  {author} {\bibfnamefont {K.}~\bibnamefont {Chen}}, \bibinfo {author}
  {\bibfnamefont {S.}~\bibnamefont {Stanescu}}, \bibinfo {author}
  {\bibfnamefont {F.}~\bibnamefont {Deschamps}}, \bibinfo {author}
  {\bibfnamefont {T.}~\bibnamefont {Moreno}}, \bibinfo {author} {\bibfnamefont
  {F.}~\bibnamefont {Polack}}, \bibinfo {author} {\bibfnamefont
  {B.}~\bibnamefont {Lagarde}}, \bibinfo {author} {\bibfnamefont {J.-P.}\
  \bibnamefont {Daguerre}}, \bibinfo {author} {\bibfnamefont {F.}~\bibnamefont
  {Marteau}}, \bibinfo {author} {\bibfnamefont {F.}~\bibnamefont {Scheurer}},
  \bibinfo {author} {\bibfnamefont {L.}~\bibnamefont {Joly}}, \bibinfo {author}
  {\bibfnamefont {J.-P.}\ \bibnamefont {Kappler}}, \bibinfo {author}
  {\bibfnamefont {B.}~\bibnamefont {Muller}}, \bibinfo {author} {\bibfnamefont
  {O.}~\bibnamefont {Bunau}}, \ and\ \bibinfo {author} {\bibfnamefont
  {P.}~\bibnamefont {Sainctavit}},\ }\href
  {http://scitation.aip.org/content/aip/journal/rsi/85/1/10.1063/1.4861191}
  {\bibfield  {journal} {\bibinfo  {journal} {Rev. Sci. Instr.}\ }\textbf
  {\bibinfo {volume} {85}},\ \bibinfo {eid} {013106} (\bibinfo {year}
  {2014})}\BibitemShut {NoStop}%
\end{thebibliography}

%

\end{document}